\documentclass[letterpaper, 10 pt, conference]{ieeeconf}  

\IEEEoverridecommandlockouts                              
\usepackage{amsmath} 
\usepackage{amssymb,amsfonts,bbm,mathrsfs,mathtools}  

\usepackage{cite}
\usepackage{algorithmic}
\usepackage{graphicx}
\usepackage{textcomp}
\usepackage{xcolor}
\usepackage{tikz}

\newtheorem{theorem}{Theorem}
\newtheorem{lemma}{Lemma}

\newtheorem{definition}{Definition}

\newtheorem{remark}{Remark}
\newtheorem{example}{Example}

\definecolor{bl}{rgb}{0,0.53,0.74}
\definecolor{red}{rgb}{0.77,0.01,0.2}
\definecolor{pur}{rgb}{0.5,0.01,0.99}
\title{\LARGE \bf
Local Observability of a Class of Feedforward Neural Networks
}

\author{Yi Yang, Victor G. Lopez, and  Matthias A. M\"uller
\thanks{This work was supported by the Deutsche Forschungsgemeinschaft (DFG, German Research Foundation) - 535860958 and the Lower Saxony Ministry for Science and Culture within the program zukunft.niedersachsen.}
\thanks{Y. Yang, V. G. Lopez, and M. A. M\"{u}ller are with the Leibniz University Hannover, Institute of Automatic Control, 30167 Hannover, Germany 
        {\tt\small \{yang, lopez, mueller\}@irt.uni-hannover.de}}%
}

\begin{document}

\maketitle
\thispagestyle{empty}
\pagestyle{empty}

\begin{abstract}

Beyond the traditional neural network training methods based on gradient descent and its variants, state estimation techniques have been proposed to determine a set of ideal weights from a control-theoretic perspective. Hence, the concept of observability becomes relevant in neural network training. In this paper, we investigate local observability of a class of two-layer feedforward neural networks~(FNNs) with rectified linear unit~(ReLU) activation functions. We analyze local observability of FNNs by evaluating an observability rank condition with respect to the weight matrix and the input sequence. First, we show that, in general, the weights of FNNs are not locally observable. Then, we provide sufficient conditions on the network structures and the weights that lead to local observability. Moreover, we propose an input design approach to render the weights distinguishable and show that this input also excites other weights inside a neighborhood. Finally, we validate our results through a numerical example.
\end{abstract}

\section{INTRODUCTION}

Neural networks (NNs) are among the most popular machine learning techniques and are widely used in the field of control engineering to approximate, for example, the dynamics of unknown plants and/or control laws, leveraging their powerful universal approximation properties \cite{bonassi2022recurrent,hunt1992neural}.  In practice, the gradient descent (GD) and the stochastic gradient decent (SGD) techniques have become the most commonly used methods for NNs training \cite{schmidhuber2015deep}. However, theoretical guarantees for convergence to the optimal weights are rarely studied \cite{li2017convergence}. 

Various training algorithms and modifications have been proposed to enhance the convergence performance and mitigate the risk of local minima \cite{li2017convergence, kleinberg2018alternative}. Beyond GD-based methods, we are particularly interested in state estimation-based procedures, where the weights of an NN are treated as the states of a nonlinear system.  In \cite{singhal1988training}, an extend Kalman filtering-based training method was proposed to incrementally update the NN weights, improving the convergence speed in practical examples. Furthermore, in \cite{bemporad2022recurrent}, this algorithm was applied to learning general recurrent neural networks and showed a competitive performance with respect to SGD in a nonlinear identification problem. Another online adaptation approach for recurrent neural networks was proposed in \cite{bonassi2022towards}, tuning the weights and deriving convergence conditions based on moving horizon estimation~(MHE). An implicit assumption of the aforementioned state estimation methods is that the states (i.e., the weights) of the system must be observable or at least detectable. However, it is well known that the optimal weights of an NN are in general not unique \cite{kleinberg2018alternative}, which implies that global observability does not hold for this system. These challenges motivate us instead to investigate  a weaker observability concept, namely local observability, for analyzing the weights in NNs. 

Local observability is a fundamental property of nonlinear systems, distinguishing a state from other states in its neighborhood. The cornerstone work was established in \cite{hermann1977nonlinear}, where various notions of local observability and controllability were introduced and analyzed theoretically for continuous-time systems.  These notions were later extended to discrete-time systems in \cite{nijmeijer1982observability} and \cite{albertini1996remarks}, and the behavior under a sampling-based scheme was analyzed in \cite{sontag1984concept}.  However, the analysis of local observability heavily depends on the traditional Lie algebraic approach, which is difficult to apply to many systems. In \cite{5400067}, a numerical method for nonlinear observability was proposed based on the empirical observability Gramian and the degree of the observability was measured. In \cite{7403218}, the authors extended the empirical observability Gramian to the case of systems with inputs. Unlike linear systems, specific control inputs must be applied to some nonlinear systems to distinguish one state from another (which are said to be persistently exciting inputs) and, therefore, it is crucial to take them into account when checking the observability rank condition. However, finding a persistently exciting (PE) input sequence for a locally observable state is nontrivial. 

In this paper, we consider two-layer FNNs with the commonly used ReLU activation functions. While \cite{vanelli2024local} proved that feedforward networks with analytic activation functions are generically locally observable (i.e., for almost all weights in a neighborhood), we investigate local observability of $\emph{all}$ weights in the neighborhood. This is important to guarantee the convergence of NN training using state estimation methods. First, we show that in general, the weights of such an NN are not necessarily locally observable. Then, we analyze a specific FNN structure that allows the satisfaction of the sufficient conditions for local observability. In particular, we show that when the number of inputs equals the number of nodes in the hidden layer, the weights forming a full rank matrix are locally observable. Moreover, we develop a method to design a set of PE  input sequences which guarantee that a given locally observable state can be distinguished from its neighbors. Although the PE input is designed with respect to a specific state, we derive a neighborhood on which all other states are rendered distinguishable by the same PE input, enabling the deduction of their local observability properties.  The results obtained in this paper are crucial for the analysis of the use of state estimation techniques, such as, e.g., MHE, as a training method for FNNs.

The rest of this paper is organized as follows. Some preliminary results on FNNs and nonlinear systems are introduced in Section \ref{section-preliminary}. A motivating example is presented to show that in general, the weights of an FNN are not necessarily locally observable in Section \ref{section-local observability}. Then, a sufficient condition for local observability of a class of FNNs is developed and a persistently exciting input design method is proposed. In Section \ref{section-numerical results}, numerical examples are presented.

\section{PRELIMINARIES AND SETUP}\label{section-preliminary}
We denote the set of integers between $a$ and $b$ by $\mathbb{Z}_{[a,b]}$, and the set of positive integers by $\mathbb{Z}^{+}$. The sequence $\{x_a,x_{a+1},\ldots,x_b\}$ is denoted by $x_{[a,b]}$. Denote the identity matrix of dimension $n$ by $\mathrm{I}_n$,  the all ones vector in $\mathbb{R}^n$ by $\mathbf{1}_n$, and the all ones matrix in $\mathbb{R}^{m\times n}$ by $\mathbf{1}_{m\times n}$. Let $\otimes$ denote the Kronecker product between two matrices of arbitrary dimensions, and $\circ$ the element-wise product between two matrices of the same dimensions. For a vector $x\in \mathbb{R}^n$, we use $\mathrm{diag}(x)$ to denote the diagonal matrix whose main diagonal consists of the elements of $x$. 
\subsection{Local Observability in Nonlinear Systems}
In the following, we introduce some definitions about local observability in regard to a nonlinear system of the form: 
\begin{subequations}\label{2-eq-nonlinearplant}
	\begin{align}
		x_{t+1}&=f(x_{t},u_{t}),\\
		y_{t}&=h(x_{t},u_{t}),\label{2-eq-nonlinearplant-2}
	\end{align}
\end{subequations}where $t\in\mathbb{Z}^+$, $x_t\in \mathbb{R}^n$, $u_t\in \mathbb{R}^m$, and $y_t\in \mathbb{R}^p$.

\begin{definition}\label{def-local observable}
	System~(\ref{2-eq-nonlinearplant}) is said to be $k$-observable at state $x$ if there exists a neighborhood $\mathcal{M}$ of $x$ and an input sequence $u_{[1,k]}\in (\mathbb{R}^m)^k$ for some $k\geq 1$ such that, for any $x^{\prime}\in \mathcal{M}$ with $x^{\prime}\neq x$, the output sequences resulting from system~(\ref{2-eq-nonlinearplant}) with initial conditions $x$ and $x^{\prime}$ are distinguishable.
\end{definition}
A similar definition is referred to as strongly locally observable when considering $k=n$ in \cite{nijmeijer1982observability}. Here, we consider a more general positive constant $k$. 

\begin{definition}
	A sequence of inputs $u_{[1,k]}$ is said to be persistently exciting for $x$ if system~(\ref{2-eq-nonlinearplant}) is $k$-observable at $x$ under $u_{[1,k]}$.
\end{definition}
\begin{definition}
	Define the $k$-observability mapping $\mathscr{H}_k$: $ \mathbb{R}^n\times (\mathbb{R}^m)^k\rightarrow(\mathbb{R}^{p})^k$ at state $x$ by
	\begin{align}
		\mathscr{H}_k(x,u_{[1,k]})=\begin{bmatrix}
			h(x,u_1)\\
			h(f(x,u_1),u_2)\\
			\vdots\\
			h(f(\cdots f(f(x,u_1),u_2),\cdots,u_k))
		\end{bmatrix},
	\end{align} for any input sequence $u_{[1,k]}$ with some $k\geq 1$. 
\end{definition}

For brevity, we omit explicitly writing the input when referring to $\mathscr{H}_k(x,u_{[1,k]})$ in the following sections. 
System~(\ref{2-eq-nonlinearplant}) is said to satisfy the \textit{observability rank condition} \cite{hermann1977nonlinear} at $x$ if the Jacobian matrix $\mathrm{d}\mathscr{H}_k(x)$ has full column rank, i.e.,
\begin{align}\label{2-eq-ORC}
	\mathrm{rank}~\mathrm{d}\mathscr{H}_k(x)=n.
\end{align} Note that the Jacobian of the $k$-observability mapping, $\mathrm{d}\mathscr{H}_k(x)$, describes how the mapping changes with respect to the state, which ensures the mapping is injective locally around $x$ if the observability rank condition holds at $x$. With the definitions above, we introduce the following lemma.
\begin{lemma}[\!\!\cite{nijmeijer1982observability}]\label{2-lem-ORC}
	If system~(\ref{2-eq-nonlinearplant}) satisfies the observability rank condition at $x$, then it is $k$-observable at $x$.
\end{lemma} 
Lemma \ref{2-lem-ORC} provides a sufficient condition for $k$-observability by exploring the first-order derivative of the mapping. This lemma was originally proposed in \cite{hermann1977nonlinear} in the continuous-time setting, extended to autonomous discrete-time systems in \cite{nijmeijer1982observability}, and later proved in \cite{albertini1996remarks} for more general nonlinear systems.  For notational simplicity, when we refer to the system as being locally observable in the following sections, we mean it is $k$-observable. Similarly, we use `observability mapping' as shorthand for `$k$-observability mapping'.
\subsection{Two-layer Feedforward Neural Networks}
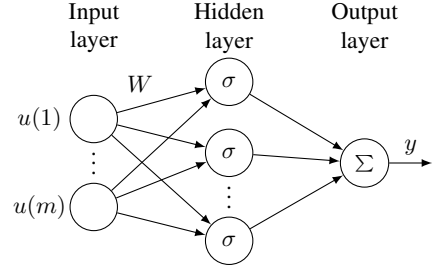
\begin{figure}[!tb]
	\centering
	\begin{tikzpicture}[scale=0.9, 
		auto, 
		every node/.style={transform shape}, 
	 every text node part/.style={align=center},
		neuron/.style={circle, draw, minimum size=0.7cm}, 
		connect/.style={draw, -latex},
		layer/.style={align=center},
		midpoint/.style = {coordinate}
		]
		\node[neuron] (I1) at (0, 0.65) {}; 
		\node at ([xshift=-0.8cm] I1) {$u(1)$};
		\node[neuron] (I2) at (0, -0.65) {}; 
		\node at ([xshift=-0.8cm] I2) {$u(m)$};
		
		\node at (0, 0.1) {\vdots}; 
		
		\node[neuron] (H1) at (2, 1.2) {$\sigma$}; 
		\node[neuron] (H2) at (2, 0.15) {$\sigma$}; 
		\node at (2, -0.4) {\vdots}; 
		\node[neuron] (H3) at (2, -1.15) {$\sigma$}; 
		
		\node[neuron] (O1) at (4, 0) {$\scriptstyle \sum$};
		
		\foreach \j in {2,3} {
			\foreach \i in {1,2} {
				\draw[connect] (I\i) -- (H\j);
			}
		}
		\draw[connect] (I1) -- node{$W$}(H1);
		\draw[connect] (I2) -- (H1);
		\foreach \i in {1,2,3} {
			\draw[connect] (H\i) -- (O1);
		}
		\node [midpoint, right of=O1, node distance=1cm] (output) {};
		\draw[connect] (O1) -- node{$y$}(output);
		\node[layer, above of=I1, node distance=1.35cm] (input) {Input \\ layer};
		\node[layer, above of=H1, node distance=0.8cm] (hidden) {Hidden \\ layer};
		\node[layer, above of=O1, node distance=2cm] (output) {Output \\ layer};
	\end{tikzpicture}
	\caption{Structure of FNNs}
	\label{fig:FNN}
\end{figure}
In this paper, we consider two-layer FNNs shown in Fig.~\ref{fig:FNN}, with $m$ nodes in the input layer, $n$ nodes in the hidden layer and one node in the output layer.\footnote{As mentioned later, the weights from the hidden layer to the output layer are all fixed to $1$. In this case, having multiple outputs would imply that all of them yield identical values and, thus, in this paper we focus on the single-output case.} The weight from the $i$th node of the input layer to the $j$th node of the hidden layer is denoted by $w_{i,j}$, the weights from the whole input layer to the $j$th node of the hidden layer are denoted by $w_j\coloneqq[w_{1,j},w_{2,j},\ldots,w_{m,j}]^{\top}$, and $W\coloneqq[w_1,w_2, \ldots,w_n]\in \mathbb{R}^{m\times n}$ is the FNN weight matrix. The activation functions in the hidden layer are identical and denoted by $\sigma$. The weights from the hidden layer to the output layer are all $1$, and the activation function in the output layer is the identity function. Selecting this specific FNN architecture simplifies our arguments in the following section, where we show that even in this case it is nontrivial to guarantee local observability. In particular, including the output weights as part of the state to be estimated further restricts the conditions for local observability, and we leave the exploration of this setting for future work.  

Based on the described FNNs setting, the function between the input $u\in\mathbb{R}^m$ and the output $y$ is
\begin{align}\label{2-eq:FNN}
	h(W,u)=\sum_{j=1}^{n}\sigma(w_j^{\top}u).
\end{align} 

For the hidden layer, we use the ReLU activation function, defined as $\sigma(a)=\max(0,a)$. Following the definition in~\cite{nair2010rectified}, $\mathrm{d}\sigma(a)$ is $0$ for $a\leq0$ and $1$ for $a>0$. 

In the following, we denote the concatenation of the weights $w_j$, $j\in \mathbb{Z}_{[1,n]}$, by $w\coloneqq (w_1^{T},w_2^{T},\ldots,w_n^{T})^{\top}\in\mathbb{R}^{mn}$. We define the indicator function $\chi(a)$ as
\begin{align}
	\chi(a) =
	\begin{cases}
		1, & a > 0, \\
		0, & a \leq 0.
	\end{cases}
\end{align}
The indicator function can be extended to vectors or matrices by applying it element-wise. Then, given an input $u_i\in \mathbb{R}^m$, and using the fact that the activation functions $\sigma$ are the ReLU functions, we have
\begin{align}\label{2-eq-derivative}
	\frac{\partial y}{\partial w}&=\left[ \frac{\partial h}{\partial w_1}, ,\ldots ,\frac{\partial h}{\partial w_n}\right]=\left[ u_i^{\top} \chi_{i,1},  \ldots ,u_i^{\top} \chi_{i,n}\right],
\end{align}where 
\begin{align}\label{2-eq-indicator}
	\chi_{i,j}\coloneqq \chi(w_j^{\top}u_i),~j\in \mathbb{Z}_{[1,n]}.
\end{align}

\section{LOCAL OBSERVABILITY ANALYSIS OF FNNS}\label{section-local observability}
In this section, we first illustrate with an example that the weights of an FNN, in general, cannot be uniquely determined from output measurements regardless of the applied inputs. Then, we consider a particular class of FNNs and investigate its local observability using the observability rank condition in (\ref{2-eq-ORC}).
\subsection{Motivating Example and Problem Formulation}
\begin{example}
	Consider the FNN in (\ref{2-eq:FNN}) with $m=1$ and $n$~$=$~$3$.   All the weights are assumed to be nonzero, since otherwise we could consider a smaller, equivalent network. In this setting, there are two possible cases for the weights $W$. First, suppose that the weights $W=[a,b,c]$ have all the same sign. Then, the indicator function $\chi(uW)$ takes values in the set $\{(1,1,1),(0,0,0)\}$ for any input $u\in \mathbb{R}$. The corresponding output sequence can only take values of either $0$ or $(a+b+c)\vert u\vert$, and therefore the individual weights cannot be determined under any input sequence. The remaining case is that two of the constants $a,b,c$ have the same sign. Hence, suppose that $a,b>0$ and $c<0$. In this case, any input $u<0$ yields the output $y=cu$, while $u>0$ yields $y=(a+b)u$. Again, the individual weights $\{a,b,c\}$ cannot uniquely be determined. Using similar arguments, it can be seen that any FNN with $m=1$ and $n>3$ inevitably has indistinguishable weights as well, violating the condition for local observability.
\end{example}

The above example shows that local observability of the weights of  FNNs is limited, as even a simple FNN structure fails to satisfy this property. Beyond the FNN structure, the observability of a particular  FNN  configuration also depends on its weights, which determine whether an input sequence exists such that the observability rank condition (\ref{2-eq-ORC}) holds, as we show in the next section.

To formally discuss the observability properties of FNNs, we must first represent it with a dynamic equation. Notice that the input-output mapping in~(\ref{2-eq:FNN}) can be represented as a dynamic system
\begin{subequations}\label{2-eq-FNN problem}
	\begin{align}
		w_{t+1}&=w_{t},\label{2-eq-FNN state}\\
		y_t&=h(w_{t},u_{t}),\label{2-eq-FNN output}
	\end{align}
\end{subequations}with $t\in \mathbb{Z}^+$, where $w_t\in \mathbb{R}^{nm}$, $u_t\in \mathbb{R}^{m}$, and $y_t\in \mathbb{R}$ are the weights (also the states), the input, and the output of FNNs at time step $t$. Variations of the model in~(\ref{2-eq-FNN problem}) have been used to train NNs by means of state estimation techniques \cite{bemporad2022recurrent, bonassi2022towards, singhal1988training}. With a slight abuse of notation, we use $h(w_{t},u_{t})$ to denote the FNN input-output mapping, where $w_t$ is the concatenation of weights at time step $t$. Notice that, since the ideal weights of an FNN are constant, the corresponding dynamic representation (\ref{2-eq-FNN state}) is static.

The output in~(\ref{2-eq-FNN problem}) is determined by the current weights and input, without influence from past inputs. The observability mapping for the FNN at state $w$ under a sequence of inputs $u_{[1,N]}$ is given by 
\begin{align}
	\mathscr{H}_N(w)=[h(w,u_1),h(w,u_2), \ldots, h(w,u_N)]^{\top}.
\end{align}Then, combining with (\ref{2-eq-derivative}) and (\ref{2-eq-indicator}), we obtain the Jacobian as
\begin{align}
	\mathrm{d}\mathscr{H}_{N}(w)&\!=\!\begin{bmatrix}
		\frac{\partial h}{\partial w_1}(w,u_1) &\frac{\partial h}{\partial w_2}(w,u_1) &\!\!\cdots\!\! &\frac{\partial h}{\partial w_n}(w,u_1)\\
		\frac{\partial h}{\partial w_1}(w,u_2) &\frac{\partial h}{\partial w_2}(w,u_2) &\!\!\cdots \!\!&\frac{\partial h}{\partial w_n}(w,u_2)\\
		\vdots&\vdots &\!\!\ddots\!\!&\vdots\\
		\frac{\partial h}{\partial w_1}(w,u_{N}) &\frac{\partial h}{\partial w_2}(w,u_{N}) &\!\!\cdots\!\! &\frac{\partial h}{\partial w_n}(w,u_{N})\\
	\end{bmatrix}\notag\\
	&\!=\!\begin{bmatrix}\label{2-eq-H-derivative}
		u_1^{\top} \chi_{1,1} & u_1^{\top} \chi_{1,2} & \cdots &u_1^{\top} \chi_{1,n}\\
		u_2^{\top} \chi_{2,1} & u_2^{\top} \chi_{2,2} & \cdots &u_2^{\top} \chi_{2,n}\\
		\vdots&\vdots&\ddots&\vdots\\
		u_{N}^{\top} \chi_{N,1} & u_{N}^{\top} \chi_{N,2} & \cdots &u_{N}^{\top} \chi_{N,n}\\
	\end{bmatrix}.
\end{align}

The Jacobian matrix $\mathrm{d}\mathscr{H}_N(w)$ can be explicitly represented as the product of a matrix formed by the input sequences  and a matrix of the indicator functions as follows, 
\begin{align}\label{2-eq-dH-factorization}
	\mathrm{d}\mathscr{H}_N(w)=\underbrace{
		\begin{bsmallmatrix}
			u_1^{\top}&~&~&~\\
			~&u_2^{\top}&~&~\\
			~&~&\ddots&~\\
			~&~&~&u_N^{\top}
	\end{bsmallmatrix}}_{T_u}\left(\underbrace{\begin{bsmallmatrix}
			\chi_{1,1} &  \chi_{1,2} & \cdots & \chi_{1,n}\\
			\chi_{2,1} &  \chi_{2,2} & \cdots & \chi_{2,n}\\
			\vdots&\vdots&\ddots&\vdots\\
			\chi_{N,1} &  \chi_{N,2} & \cdots & \chi_{N,n}
	\end{bsmallmatrix}}_{T_{\chi}}\otimes\mathrm{I}_{m}\right),
\end{align}where $T_u\in\mathbb{R}^{N\times Nm}$, $T_{\chi}\in\mathbb{R}^{N\times n}$. The indicator function of $u_i^{\top}W$ can be expressed as $\chi(u_i^{\top}W)=\left[ \chi(u_i^Tw_1), \chi(u_i^Tw_2), \ldots ,\chi(u_i^Tw_n)\right]$. Recalling the definition of $\chi_{i,j}$ in~(\ref{2-eq-indicator}), and that $w_j^{\top}u_i=u_i^{\top}w_j$ is a scalar, yields $\chi(u_i^{\top}W)=\left[ \chi_{i,1}, \chi_{i,2}, \ldots ,\chi_{i,n}\right]$. Denoting $U\coloneqq[u_1,\ldots,u_N]^{\top}\in \mathbb{R}^{N\times m}$, it follows that $T_{\chi}=\chi(UW)$

This factorization allows to determine some conditions for the full rank requirement of $\mathrm{d}\mathscr{H}_N(w)$. For example, consider a full row rank weight matrix $W\in \mathbb{R}^{m\times n}$ with $m<n$. Then, without loss of generality\footnote{Note that permuting the columns of W corresponds to permuting the order of the nodes in the hidden layer of FNNs.}, we can write $T_{\chi}=\chi(U[W_1,W_2])$, where $W_1\in\mathbb{R}^{m\times m}$ is non-singular, and $W_2\in \mathbb{R}^{m \times (n-m)}$ is a linear combination of the columns of $W_1$. A necessary condition to ensure that $\mathrm{d}\mathscr{H}_N(w)$ has full column rank is that $T_{\chi}$ has full column rank. However, the dependence between $W_1$ and $W_2$ restricts the potential results of $\chi(UW_2)$, which imposes further constraints on $W$.  

In the next section, we show that specific FNN architectures allow their weights to be locally observable. In particular, we consider the class of FNNs with the same number of inputs as nodes in the hidden layer, i.e., $m=n$, and provide a sufficient condition for a locally observable state.

\subsection{A Sufficient Condition for a Locally Observable State }
The following result presents a sufficient condition for a locally observable state in (\ref{2-eq-FNN problem}). For the clarity of our presentation, we show this by constructing a specific persistently exciting input sequence. Later, in the next subsection, we develop a method to design more general persistently exciting inputs. 
\begin{theorem}\label{th-local observable}
	Consider the FNN in~(\ref{2-eq-FNN problem}) with $m=n$. The state $w$ is locally observable if the corresponding weight matrix $W$ is non-singular.
\end{theorem}
\begin{proof}
	We show that there exists an input sequence $u_{[1,N]}$, under which the observability rank condition~(\ref{2-eq-ORC}) is satisfied at the state $w$. In particular, consider the case $N=n^{2}$, where $n$ is the number of nodes in the hidden layer (also the number of nodes in the input layer due to $m=n$) and $N$ is the length of the input sequence. Denote again $U\coloneqq[u_1,u_2,\ldots,u_N]^{\top}$, and a potential input is constructed as $U=BW^{-1}$, where $B\coloneqq [B_1^{\top}, B_2^{\top},\ldots,B_n^{\top}]^{\top}$ is a constant matrix in $\mathbb{R}^{N\times n}$. The block element $ B_k\in \mathbb{R}^{n\times n}, k\in \mathbb{Z}_{[1,n]}$ is defined as a matrix where the $k$th column is all $1$, the diagonal elements are all $-1$ except for the $k$th column, and the rest of the elements are $0$, e.g.,
	\begin{align*}
		B_1\coloneqq 	\begin{bmatrix}
			b_1 \\b_2 \\ \vdots\\ b_n
		\end{bmatrix}=\begin{bmatrix}
			1 &0 &\cdots &0\\
			1 &-1&\cdots&0	\\
			\vdots&\ddots&\ddots&\vdots\\
			1&0&\cdots&-1
		\end{bmatrix}.
	\end{align*}
	
	From the definition of $B_k$, the indicator function $\chi(B_k)$ yields a matrix with the elements in the $k$th column equal to $1$ and the rest of the elements are $0$. This can be expressed as $\chi(B_k)=e_k^{\top}\otimes \mathbf{1}_n$, where $e_k\in\mathbb{R}^n$ is the standard basis vector with all elements equal to zero except for the $k$th element, which is  $1$.  From the construction of the input sequence, we have $UW=B$. Therefore, it follows that $\chi(UW)=\chi(B)=\mathrm{I}_n\otimes \mathbf{1}_n$ and $u_i^{\top}W=b_i$.
	
	Combining the results that $\left[ \chi_{i,1}, \chi_{i,2}, \ldots ,\chi_{i,n}\right]=\chi(u_i^{\top}W)=\chi(b_i)$ and $\chi(B)=\mathrm{I}_n\otimes \mathbf{1}_n$, the Jacobian matrix of the observability mapping in~(\ref{2-eq-H-derivative}) is block diagonal given by
	\begin{align}\label{2-eq-block}
		&\mathrm{d}\mathscr{H}_N(w)=\begin{bmatrix}
			H_1 & \mathbf{0} & \cdots &\mathbf{0}\\
			\mathbf{0}  & H_2 & \cdots &\mathbf{0}\\
			\vdots&\vdots&\ddots&\vdots\\
			\mathbf{0} & \mathbf{0} & \cdots &H_n\\
		\end{bmatrix},\\
		&H_k\!=\!\begin{bmatrix}
			u_{(k-1)n+1}, u_{(k-1)n+2}, \cdots, u_{kn}
		\end{bmatrix}^{\top}, k\in \mathbb{Z}_{[1,n]}.\label{3-eq-Hk}
	\end{align}Since $W$ is non-singular and $B_k$ is constructed as a non-singular matrix, we have $\mathrm{rank}(H_k)=\mathrm{rank}(B_kW^{-1})=n$. Therefore, $H_k$ is non-singular, which ensures  that $\mathrm{d}\mathscr{H}_{N}(w)$  has full column rank. This shows that there exists an input sequence such that system~(\ref{2-eq-FNN problem}) satisfies the observability rank condition at state $w$.  
\end{proof}
\begin{remark}
	Theorem~\ref{th-local observable} provides a sufficient condition to identify a locally observable state by using the observability rank condition.The minimal length of the input sequences for this condition to hold is $N=n^2$. This is because the number of columns in $\mathrm{d}\mathscr{H}_N(w)$ is $n^2$. This means that, while longer persistently exciting input sequences (with $N>n^2$) exist, any locally observable state satisfying the condition in Theorem~\ref{th-local observable} can be excited by an input sequence of minimal length~$n^2$ as shown in the proof of Theorem~\ref{th-local observable}.    
\end{remark}

\subsection{Persistently Exciting Input Design}\label{section-input design}
In this section, we propose an approach to design a set of persistently exciting input sequences for a given locally observable state $w$ for the described FNN setting in (\ref{2-eq:FNN}). This requires the satisfaction of the observability rank condition at~$w$. Compared to Theorem~\ref{th-local observable}, this results in a more general set of input sequences of length $N=n^2$ that can excite a locally observable state.
\begin{theorem}\label{th-orc}
	Consider the FNN in~(\ref{2-eq-FNN problem}) with $m=n$. Assume the weight matrix $W$ corresponding to the state $w$ is non-singular. For any $B=[B_1^{\top},B_2^{\top},\ldots,B_n^{\top}]^{\top}$, $B_k^{\top}\in\mathbb{R}^{n\times n}$, $k\in\mathbb{Z}_{[1,n]}$, satisfying
	\begin{align}
		&\mathrm{rank}(B_k)=n,k\in\mathbb{Z}_{[1,n]},\label{4-th-c1}\\
		&\chi(B)=T\otimes \mathbf{1}_{n},\label{4-th-c2}
	\end{align}where $T\in\mathbb{R}^{n\times n}$ is any non-singular matrix whose elements are either $1$ or $0$, the input sequence $u_{[1,N]}$ derived by
	\begin{align}
		U\coloneqq [u_1, u_2,\ldots,u_N]^{\top}= BW^{-1},\label{4-th-c3}
	\end{align} is persistently exciting.
\end{theorem} 
\begin{proof}
	We prove the theorem by showing that any input derived from (\ref{4-th-c3}) ensures that the observability rank condition holds. To this end, for contradiction, assume that $\textrm{rank}(\mathrm{d}\mathscr{H}_N(w))<n^2$. Then, there exist some coefficients $c_{1},\ldots, c_{n^2}$ which are not all zero, such that with (\ref{2-eq-H-derivative}), we~have
	\begin{align}
		\begin{bmatrix}
			u_1^{\top} \chi_{1,1} & u_1^{\top} \chi_{1,2} & \cdots &u_1^{\top} \chi_{1,n}\\
			u_2^{\top} \chi_{2,1} & u_2^{\top} \chi_{2,2} & \cdots &u_2^{\top} \chi_{2,n}\\
			\vdots&\vdots&\ddots&\vdots\\
			u_{N}^{\top} \chi_{N,1} & u_{N}^{\top} \chi_{N,2} & \cdots &u_{N}^{\top} \chi_{N,n}\\
		\end{bmatrix}\begin{bmatrix}
			c_1\\c_2\\ \vdots \\ c_{n^2}
		\end{bmatrix}=\mathbf{0}.\label{4-eq-th-1}
	\end{align}By taking $n$-row subsets of~(\ref{4-eq-th-1}), we have
	\begin{align}
		\begin{bmatrix}\label{4-eq-th-subset rows}
			u_{kn+1}^{\top} \chi_{kn+1,1}  & \cdots &u_{kn+1}^{\top} \chi_{kn+1,n}\\
			u_{kn+2}^{\top} \chi_{kn+2,1}  & \cdots &u_{kn+2}^{\top} \chi_{kn+2,n}\\
			\vdots&\ddots&\vdots\\
			u_{(k+1)n}^{\top} \chi_{(k+1)n,1} &  \cdots &u_{(k+1)n}^{\top} \chi_{(k+1)n,n}\\
		\end{bmatrix}\begin{bmatrix}
			c_1\\c_2\\ \vdots \\ c_{n^2}
		\end{bmatrix}=\mathbf{0},
	\end{align}where $ k\in\mathbb{Z}_{[0,n-1]}$. 
	
	Recalling the definition of $T_{\chi}$ in (\ref{2-eq-dH-factorization}), from~(\ref{4-th-c3}), it is clear that $T_{\chi}=\chi(UW)=\chi(B)$. Combining this with~(\ref{4-th-c2}), we have
	\begin{align}\label{4-eq-th-subset indicator}
		\begin{bmatrix}
			\chi_{kn+1,1} &   \cdots & \chi_{kn+1,n}\\
			\chi_{kn+2,1} &  \cdots & \chi_{kn+2,n}\\
			\vdots&\ddots&\vdots\\
			\chi_{(k+1)n,1} &   \cdots & \chi_{(k+1)n,n}
		\end{bmatrix}=T_{k+1}\otimes\mathbf{1}_n,
	\end{align}where $T_{k+1}$ is the $(k+1)$th row of $T$. Since $T$ is required to be non-singular,  $T_{k+1}$ has at least one element equal to $1$. Therefore, the columns of the left-hand side matrix in~(\ref{4-eq-th-subset indicator}) are either all ones or zeros vectors, and at least one column is all ones vector. This results in the left-most matrix in~(\ref{4-eq-th-subset rows}) being composed of blocks, each of which is either $H_{k+1}$ as defined in~(\ref{3-eq-Hk}) or $\mathbf{0}_{n\times n}$.
	
	Combining with~(\ref{4-th-c3}), we can obtain that $H_k=B_kW^{-1}$, for $k\in \mathbb{Z}_{[1,n]}$, is non-singular, as condition~(\ref{4-th-c1}) guarantees $B_k$ is non-singular. Therefore, the only way to ensure the linear dependence among columns of the left-most matrix in~(\ref{4-eq-th-subset rows}) is that the $j\mathrm{th},(n+j)\mathrm{th},\ldots,((n-1)n+j)\mathrm{th}$ columns of this matrix are linearly dependent for all $j\in \mathbb{Z}_{[1,n]}$, i.e., we have
	\begin{align}\label{4-eq-coefficient0}
		\mathrm{diag}(u_{kn+i})\underbrace{\begin{bmatrix}
				c_1&c_{n+1}&\cdots&c_{n(n-1)+1}\\
				c_2&c_{n+2}&\cdots&c_{n(n-1)+2}\\
				\vdots&\vdots&\ddots&\vdots\\
				c_{n}&c_{2n}&\cdots&c_{n^2}\\
		\end{bmatrix}}_{C}T_{k+1}^{\top}=\mathbf{0},
	\end{align}
	for all $i\!\in\!\mathbb{Z}_{[1,n]}$ and $k\!\in\!\mathbb{Z}_{[0,n-1]}$, which can be represented as 
	\begin{align}
		\mathrm{diag}(C T_{k+1}^{\top})u_{kn+i}=\mathbf{0},
	\end{align}for all $i\in \mathbb{Z}_{[1,n]}$ and $k\in\mathbb{Z}_{[0,n-1]}$. Therefore, we obtain
	\begin{align}
		\mathrm{diag}(C T_{k+1}^{\top}) H_{k+1}^{\top}=\mathbf{0}, k\in\mathbb{Z}_{[0,n-1]}.
	\end{align}
	Since $H_{k+1}$ is non-singular, it follows that $C T_{k+1}^{\top}=\mathbf{0}$, $ k\in\mathbb{Z}_{[0,n-1]}$. Combining for all $k\in \mathbb{Z}_{[0,n-1]}$, we obtain $C\begin{bmatrix}
		T_1^{\top},T_2^{\top},\ldots,T_n^{\top}
	\end{bmatrix}=\mathbf{0}$, which leads to $c_{1}=\cdots= c_{n^2}=0$, since $T$ is non-singular. However, this conclusion contradicts our initial assumption in~(\ref{4-eq-th-1}). Therefore, the observability mapping $\mathrm{d}\mathscr{H}_N(w)$ has full column rank and the observability rank condition holds at $w$ under the designed input  $U$ if conditions~(\ref{4-th-c1}) and~(\ref{4-th-c2}) are satisfied.
\end{proof}
Theorem~\ref{th-orc} provides an approach to design persistently exciting input sequences. A close inspection of the proof of Theorem~\ref{th-local observable} shows that the input designed there is, in fact, a special case of the more general set of input sequences considered in Theorem~\ref{th-orc} (with $T=\mathrm{I}_n$ and a specific choice of $B$). 

The inputs designed in Theorem~\ref{th-orc} require prior knowledge of a particular state with non-singular weight matrix $W$. In the following theorem, we show that if an input is designed to make a state $w$ distinguishable, then other states in a neighborhood around $w$ will also be rendered distinguishable by the same input. 
\begin{theorem}\label{th-observable set}
	Consider a locally observable state $w$ of system~(\ref{2-eq-FNN problem}) with a corresponding non-singular weight matrix $W$, and a persistently exciting input $U=BW^{-1}$ with $B$ as in Theorem~\ref{th-orc}. The state $w^{\prime}$ with corresponding weight matrix $W^{\prime}$ is also locally observable and excited by $U$ if there exists a matrix $K\in\mathbb{R}^{N\times n}$ with all elements in $(-1,+\infty)$, such that 
	\begin{align}\label{th-observable set-c1}
		U\delta=K\circ UW
	\end{align}holds, where $\delta=W^{\prime}-W$.
\end{theorem} 
\begin{proof}
	Denote $B=UW$  and $B^{\prime}=UW^{\prime}$. We obtain $B^{\prime}=U(W+\delta)=B+U\delta$.
	We continue our proof by showing that the observability rank condition is satisfied at state $w^{\prime}$ under the input $U$.
	
	Condition~(\ref{th-observable set-c1}) establishes a relationship between elements in $U\delta$ and $UW$, where the condition on $K$ ensures that each element of $U\delta$ is given by the corresponding element in $UW$ multiplied by a constant greater than $-1$. This implies that the sign of elements in $B+U\delta$ are preserved the same as those in $B$. Therefore, we have $\chi(B+U\delta)=\chi(B)$, which ensures the satisfaction of~(\ref{4-th-c2}). Following the same procedures as in the proof of Theorem~\ref{th-orc}, we obtain that~(\ref{4-eq-th-subset rows}) and (\ref{4-eq-th-subset indicator}) hold for the state $w^{\prime}$.
	
	Furthermore, the persistently exciting input $U$ is obtained from~(\ref{4-th-c3}), which ensures $H_k, k\in \mathbb{Z}_{[1,n]}$ has full column rank. Combining with the rest of the proof for Theorem~\ref{th-orc}, the observability mapping $\mathrm{d}\mathscr{H}_N(w^{\prime})$ has full column rank. Therefore, the unknown state $w^{\prime}$ is locally observable and excited by the same input $U$.
\end{proof}

Theorem~\ref{th-observable set} imposes a constraint on the distance between the locally observable state $w$ used to design the persistently exciting input, and the unknown states $w^{\prime}$ that are also excited by it.  A corresponding set of such states $w^{\prime}$ can be determined by computing $\delta$ for arbitrary matrices $K$ satisfying the conditions of Theorem~\ref{th-observable set}. In particular, fix any such $K$, compute $\delta = U^{\dagger}(K\circ UW)$, and then verify whether this $\delta$  satisfies (\ref{th-observable set-c1}). In fact, if this is the case, then $\delta$ is the unique solution to (\ref{th-observable set-c1}), because $U^{\dagger}=(U^{\top}U)^{-1}U^{\top}$ is a left inverse since $U=BW^{-1}$ has full column rank. Note that if all elements of $B=UW$ are nonzero\footnote{Note that this not a restriction. In fact, if B has some zero elements, one can slightly perturb $B$ by setting these elements to a value slightly less than zero such that (\ref{4-th-c1}) still holds. This does not change $\chi(B)$ and hence also (\ref{4-th-c2}) is still satisfed, i.e., this modified $B$ still satisfies the conditions of Theorem~\ref{th-orc}.}, the set of weights $w^{\prime}$ constructed in this fashion always contains a non-empty neighborhood of $w$. This is the case since for $\delta$ small enough, (\ref{th-observable set-c1}) can always be satisfied by suitably choosing $K$ since the allowed interval for the elements of $K$ contains $0$ in its interior. On the other hand, also weights $w^{\prime}$ that are potentially far away from $w$ might be included in this set (as long as a matrix $K$ exists such that (\ref{th-observable set-c1}) holds, which is easy to check for a given $w^{\prime}$ and hence $\delta$), as illustrated in the example in Section~\ref{section-numerical results}.

\section{NUMERICAL RESULTS}\label{section-numerical results}
\begin{figure}[tb]	
	\centering
	\includegraphics[width=0.45\textwidth]{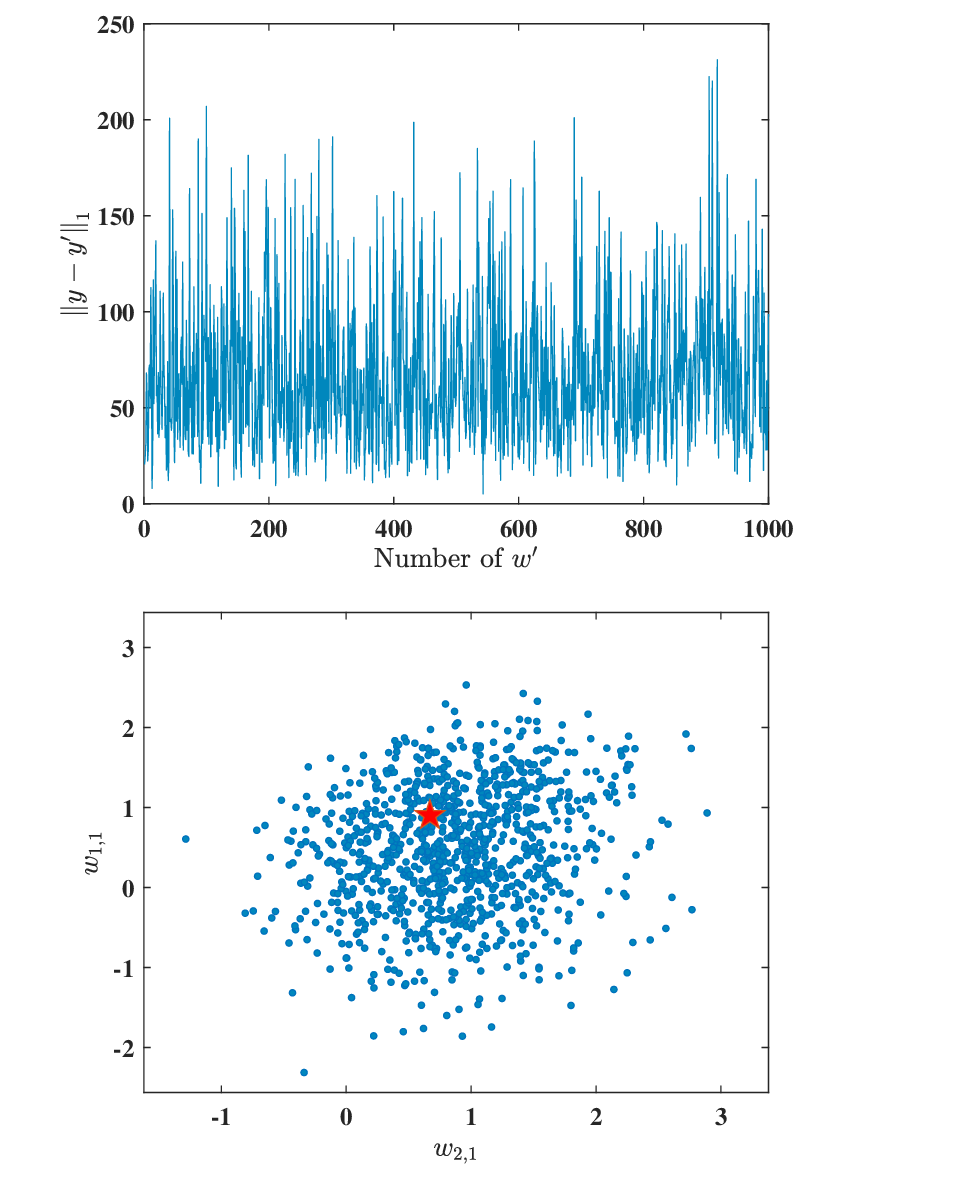}
	\vspace{-15pt}
	\caption{Top: the $\ell_1$-norm of the output difference between the sequence resulting from $w$ and $w^{\prime}$ under the same input  $U$. Bottom: the weight from the input layer to the first node in the hidden layer of $w$ (red star) and those of generated $w^{\prime}$ (blue points).}
	\label{fig-error}
\end{figure}
In this section, we demonstrate the effectiveness of the proposed input design method to guarantee local observability of an FNN with same number of nodes ($m=n=3$) in the input layer and the hidden layer. First, we design an input sequence based on Theorem~\ref{th-orc} and illustrate that the given state is distinguishable from its neighbors, which are generated using Theorem~\ref{th-observable set}. 

A prior known state $w$ is chosen randomly whose weight matrix $W$ is
\begin{align}
	W=\begin{smallmatrix}
		\begin{bmatrix}
			0.67&    0.07&0.15\\
			0.90&    0.42&0.09\\0.72&0.91&0.51
		\end{bmatrix}
	\end{smallmatrix}.
\end{align}Then, we choose a matrix $B$ that satisfies conditions~(\ref{4-th-c1}) and~(\ref{4-th-c2}) by first constructing an arbitrary non-singular $T$ as
\begin{align*}
	T=\begin{smallmatrix}
		\begin{bmatrix}
			0& 1&    0\\1&   1&1\\
			0&    0&    1
		\end{bmatrix}
	\end{smallmatrix}.
\end{align*}With the designed $B$ and weight matrix $W$, a persistently exciting input $U$ of length $N=9$ is obtained following the design method in~(\ref{4-th-c3}).  To find a neighborhood of $W$, where states can also be excited by the same input $U$, we apply the results in Theorem~\ref{th-observable set} and generate $1000$ different states $w^{\prime}$ with the elements of $K$ in $(-1,+\infty)$. In this way, the generated states are excited by the same input according to Theorem~\ref{th-observable set}. In Fig.~\ref{fig-error}, we observe that all $w^{\prime}$ can be distinguished from $w$, and that a large neighborhood of $w$ is obtained, demonstrating the effectiveness of our method.

\section{CONCLUSION}
In this paper, we formulated a class of FNNs as dynamic systems and investigated their local observability.  We showed that local observability does not necessarily hold for general FNN structures and, therefore, further research is needed to investigate the convergence of FNN training using state estimation techniques. For the particular case where the number of inputs is equal to that of nodes in the hidden layer,  local observability is possible when the corresponding weight matrix has full rank. For this case, we provided an approach to design persistently exciting input sequences and showed that such inputs not only guarantee the distinguishability of one state, but also of states in a neighborhood around it. Further research is needed to analyze local observability for more general FNN architectures.

\bibliographystyle{IEEEtran}
\bibliography{reference}
\end{document}